\documentclass[aps,showpacs,preprint]{revtex4}
\usepackage{graphicx}
\usepackage{amssymb,amsmath}

\begin{document}

\title{On the Quantum Reconstruction of 
the Riemann zeros
}

\author{Germ\'an Sierra}

\affiliation{Instituto de F\'{\i}sica Te\'orica, CSIC-UAM, Madrid, Spain}
\date{October, 2007}


\bigskip\bigskip\bigskip\bigskip

%
\font\numbers=cmss12
\font\upright=cmu10 scaled\magstep1
\def\stroke{\vrule height8pt width0.4pt depth-0.1pt}
\def\topfleck{\vrule height8pt width0.5pt depth-5.9pt}
\def\botfleck{\vrule height2pt width0.5pt depth0.1pt}
\def\Zmath{\vcenter{\hbox{\numbers\rlap{\rlap{Z}\kern
0.8pt\topfleck}\kern 2.2pt
                   \rlap Z\kern 6pt\botfleck\kern 1pt}}}
\def\Qmath{\vcenter{\hbox{\upright\rlap{\rlap{Q}\kern
                   3.8pt\stroke}\phantom{Q}}}}
\def\Nmath{\vcenter{\hbox{\upright\rlap{I}\kern 1.7pt N}}}
\def\Cmath{\vcenter{\hbox{\upright\rlap{\rlap{C}\kern
                   3.8pt\stroke}\phantom{C}}}}
\def\Rmath{\vcenter{\hbox{\upright\rlap{I}\kern 1.7pt R}}}
\def\Z{\ifmmode\Zmath\else$\Zmath$\fi}
\def\Q{\ifmmode\Qmath\else$\Qmath$\fi}
\def\N{\ifmmode\Nmath\else$\Nmath$\fi}
\def\C{\ifmmode\Cmath\else$\Cmath$\fi}
\def\R{\ifmmode\Rmath\else$\Rmath$\fi}
\def\H{{\cal H}}
\def\NN{{\cal N}}

\begin{abstract}
We discuss a possible spectral realization of the Riemann zeros
based on the Hamiltonian $H = xp$ perturbed by 
a term that depends on two potentials,  which 
are related to the Berry-Keating 
semiclassical constraints. 
We find perturbatively 
the potentials whose Jost function 
is given by the zeta function $\zeta(\sigma - i t)$ 
for $\sigma > 1$. For $\sigma = 1/2$ we find 
the potentials that yield the smooth approximation
to the zeros. 
We show that the existence
of potentials realizing the 
zeta function at $\sigma = 1/2$, 
as a Jost function,
would imply that the Riemann zeros are point
like spectrum embedded in the continuum, resolving
in that way the emission/spectral interpretation. 
\end{abstract}

\pacs{02.10.De, 05.45.Mt, 11.10.Hi}

\maketitle

\vskip 0.2cm

%
%
%
%
\def\oti{{\otimes}}
\def\lb{ \left[ }
\def\rb{ \right]  }
\def\tilde{\widetilde}
\def\bar{\overline}
\def\hat{\widehat}
\def\*{\star}
\def\[{\left[}
\def\]{\right]}
\def\({\left(}      \def\BL{\Bigr(}
\def\){\right)}     \def\BR{\Bigr)}
    \def\BBL{\lb}
    \def\BBR{\rb}
%
%
\def\zb{{\bar{z} }}
\def\zbar{{\bar{z} }}
\def\frac#1#2{{#1 \over #2}}
\def\inv#1{{1 \over #1}}
\def\half{{1 \over 2}}
\def\d{\partial}
\def\der#1{{\partial \over \partial #1}}
\def\dd#1#2{{\partial #1 \over \partial #2}}
\def\vev#1{\langle #1 \rangle}
\def\ket#1{ | #1 \rangle}
\def\rvac{\hbox{$\vert 0\rangle$}}
\def\lvac{\hbox{$\langle 0 \vert $}}
\def\2pi{\hbox{$2\pi i$}}
\def\e#1{{\rm e}^{^{\textstyle #1}}}
\def\grad#1{\,\nabla\!_{{#1}}\,}
\def\dsl{\raise.15ex\hbox{/}\kern-.57em\partial}
\def\Dsl{\,\raise.15ex\hbox{/}\mkern-.13.5mu D}
%
%
\def\ga{\gamma}     \def\Ga{\Gamma}
\def\be{\beta}
\def\al{\alpha}
\def\ep{\epsilon}
\def\vep{\varepsilon}
\def\dep{d}
\def\arc{{\rm Arctan}}
\def\la{\lambda}    \def\La{\Lambda}
\def\de{\delta}     \def\De{\Delta}
\def\om{\omega}     \def\Om{\Omega}
\def\sig{\sigma}    \def\Sig{\Sigma}
\def\vphi{\varphi}
\def\sign{{\rm sign}}
\def\he{\hat{e}}
\def\hf{\hat{f}}
\def\hg{\hat{g}}
\def\ha{\hat{a}}
\def\hb{\hat{b}}
\def\f{{\bf f}}
\def\g{{\bf g}}
\def\a{{\bf a}}
\def\b{{\bf b}}

%
%
\def\CA{{\cal A}}   \def\CB{{\cal B}}   \def\CC{{\cal C}}
\def\CD{{\cal D}}   \def\CE{{\cal E}}   \def\CF{{\cal F}}
\def\CG{{\cal G}}   \def\CH{{\cal H}}   \def\CI{{\cal J}}
\def\CJ{{\cal J}}   \def\CK{{\cal K}}   \def\CL{{\cal L}}
\def\CM{{\cal M}}   \def\CN{{\cal N}}   \def\CO{{\cal O}}
\def\CP{{\cal P}}   \def\CQ{{\cal Q}}   \def\CR{{\cal R}}
\def\CS{{\cal S}}   \def\CT{{\cal T}}   \def\CU{{\cal U}}
\def\CV{{\cal V}}   \def\CW{{\cal W}}   \def\CX{{\cal X}}
\def\CY{{\cal Y}}   \def\CZ{{\cal Z}}

\def\Hp{{\mathbb{H}^2_+}} 
\def\Hm{{\mathbb{H}^2_-}}

\def\rvac{\hbox{$\vert 0\rangle$}}
\def\lvac{\hbox{$\langle 0 \vert $}}
\def\comm#1#2{ \BBL\ #1\ ,\ #2 \BBR }
\def\2pi{\hbox{$2\pi i$}}
\def\e#1{{\rm e}^{^{\textstyle #1}}}
\def\grad#1{\,\nabla\!_{{#1}}\,}
\def\dsl{\raise.15ex\hbox{/}\kern-.57em\partial}
\def\Dsl{\,\raise.15ex\hbox{/}\mkern-.13.5mu D}
%
%
%
\font\numbers=cmss12
\font\upright=cmu10 scaled\magstep1
\def\stroke{\vrule height8pt width0.4pt depth-0.1pt}
\def\topfleck{\vrule height8pt width0.5pt depth-5.9pt}
\def\botfleck{\vrule height2pt width0.5pt depth0.1pt}
\def\Zmath{\vcenter{\hbox{\numbers\rlap{\rlap{Z}\kern
0.8pt\topfleck}\kern 2.2pt
                   \rlap Z\kern 6pt\botfleck\kern 1pt}}}
\def\Qmath{\vcenter{\hbox{\upright\rlap{\rlap{Q}\kern
                   3.8pt\stroke}\phantom{Q}}}}
\def\Nmath{\vcenter{\hbox{\upright\rlap{I}\kern 1.7pt N}}}
\def\Cmath{\vcenter{\hbox{\upright\rlap{\rlap{C}\kern
                   3.8pt\stroke}\phantom{C}}}}
\def\Rmath{\vcenter{\hbox{\upright\rlap{I}\kern 1.7pt R}}}
\def\Z{\ifmmode\Zmath\else$\Zmath$\fi}
\def\Q{\ifmmode\Qmath\else$\Qmath$\fi}
\def\N{\ifmmode\Nmath\else$\Nmath$\fi}
\def\C{\ifmmode\Cmath\else$\Cmath$\fi}
\def\R{\ifmmode\Rmath\else$\Rmath$\fi}

\def\barray{\begin{eqnarray}}
\def\earray{\end{eqnarray}}
\def\beq{\begin{equation}}
\def\eeq{\end{equation}}

\def\no{\noindent}

\def\s{\sigma}
\def\Ga{\Gamma}

\def\L{{\cal L}}
\def\g{{\bf g}}
\def\K{{\cal K}}
\def\I{{\cal I}}
\def\M{{\cal M}}
\def\F{{\cal F}}

\def\Im{{\rm Im}}
\def\Re{{\rm Re}}
\def\ti{{\tilde{\phi}}}
\def\tR{{\tilde{R}}}
\def\tS{{\tilde{S}}}
\def\tF{{\tilde{\cal F}}}

\section{Introduction}

One of the most important problems in Mathematics
is the proof of the Riemann hypothesis (RH) which
states that the non trivial zeros of the classical
zeta function all have real part equal to 1/2 
\cite{Edwards,Titchmarsh2}. 
The importance of this conjecture lies in its connection
with the prime numbers. If the RH is true then 
the statistical distribution of the primes will
be constrained in the most favorable way. According
to Michael Berry the truth of the RH  would mean that ``there is
music in the primes'' \cite{Berry1,B-chaos}. Otherwise, in  
the words of Bombieri, the
failure of the RH would create havoc in the 
distribution of the prime numbers \cite{Bombieri}.

In so far, the proof of the RH has resisted the
attempts of many and most prominent 
mathematicians and physicist for more than a century, 
which explains in  part its popularity 
\cite{Sarnak,Conrey,Watkins}.
For these and other reasons 
the RH stands as one of the most fundamental
problems in Mathematics for the XXI century with 
possible implications in Physics. In fact, physical
ideas and techniques 
could probably be essential for a proof of the RH
\cite{Rosu,Elizalde}.
This suggestion goes back to Polya and Hilbert
which, according to the standard lore, proposed
that the imaginary part of the non trivial 
Riemann zeros are the eigenvalues of a
self-adjoint operator $H$ and hence real numbers.
In the language of Quantum Mechanics
the operator $H$ would be nothing but a Hamiltonian
whose spectrum contains the Riemann zeros. 

The Polya-Hilbert conjecture was for a long time
regarded as a wild speculation until 
the works of Selberg in the 50's and those of 
Montgomery in the 70's. Selberg 
found a remarkable duality between 
the length of geodesics on a Riemann
surface and the  eigenvalues of the 
 Laplacian operator defined on it \cite{Selberg}. This 
duality is encapsulated in the so called
Selberg trace formula, which has a strong similarity
with the Riemann explicit formula
relating the zeros and the prime numbers. 
The Riemann zeros would correspond
to the eigenvalues, and the primes to the 
geodesics. This classical versus quantum 
version of the primes and the zeros was also at the heart of 
the so called  Quantum Chaos approach to the RH (see later). 
Quite independently
of the Selberg work, Montgomery showed that the
Riemann zeros are distributed randomly and 
obeying locally the statistical law of the 
Random Matrix Models (RMM) \cite{Mont}. 
The RMM were originally proposed
to explain the chaotic behaviour of the spectra of nuclei
but it has applications in another branches
of Physics, specially in  Condensed Matter \cite{Mehta}. 
There are several universality  classes
of random matrices, and it turns out that
the one related to the 
Riemann zeros is the gaussian unitary ensemble (GUE)
associated to random hermitean matrices. 
Montgomery analytical results  found an impressive
numerical confirmation in the works of
Odlyzko in the 80's, so that the GUE 
law, as applied  to the Riemann zeros 
is nowadays called the Montgomery-Odlyzko law \cite{Odl}.

It is worth to mention that 
the prime numbers, unlike the Riemann zeros, 
are distributed almost at random over
the set of integers. Indeed, it is believed
that one can find arbitrary pairs of 
nearby odd numbers $n, n+2$, as well
as pairs arbitrarily separated. The only thing
known about the distribution of the primes
is the Gauss law according to which the n$^{\rm th}$
prime $p_n$ behaves asymptotically 
as $p_n \sim n/\log n$ \cite{Edwards}. This
statement is called the Prime Number Theorem proved by 
Hadamard and de la Vallée-Poussin in 1896. 
If the RH is true then the deviation from the Gauss law
is of order $\sqrt{n} \; \log n$. The analogue of the
Gauss law for the imaginary part of the Riemann zeros
(called it $E$) is given by the Riemann law where
the  n$^{\rm th}$-zero behaves as 
$E_n \sim 2 \pi n/\log n$. Hence, large
prime numbers are progressively scarced, while large
Riemann zeros abound.

An important hint  suggested by the  Montgomery-Odlyzko law
is that the Polya-Hilbert Hamiltonian $H$ 
must break the time reversal symmetry. The reason being that
the GUE statistics describes random Hamiltonians
where this symmetry is broken. A simple example 
is provided by materials with impurities 
 subject to an external magnetic field,
as in the Quantum Hall effect.  

A further step in the Polya-Hilbert-Montgomery-Odlyzko
pathway was taken by Berry \cite{Berry1,B-chaos}. 
who noticed a similarity between the formula
yielding the  fluctuations of the number of zeros, 
around its average position $E_n \sim 2 \pi n/ \log n$,
and a formula giving the fluctuations of the
energy levels of a Hamiltonian obtained
by the quantization of a classical chaotic system \cite{Gutzwiller}. 
The comparison between these two formulas suggests
that the prime numbers $p$  correspond to the isolated
periodic orbits whose period is $\log p$. In the
Quantum Chaos scenario the prime numbers appear
as classical objects, while the Riemann zeros are
quantal. This classical/quantum interpretation
of the primes/zeros is certainly reminiscent
of the one underlying the Selberg trace formula
mentioned earlier. One success of the 
Quantum Chaos approach is that it explains the
deviations from the GUE law of the zeros found
numerically by Odlyzko. The similarity between
the fluctuation formulas 
described above, while rather appealing, has
a serious drawback observed by Connes which
has to do with an overall sign difference between them \cite{Connes}. 
It is as if the periodic orbits were missing 
in the underlying classical chaotic dynamics, a fact that
is difficult to understand physically. This and other
observations lead  Connes to propose a rather abstract
approach to the RH based on discrete mathematical
objects known as adeles \cite{Connes}. 
The final outcome of Connes
work is a trace formula whose proof, not yet found, 
amounts to that of a generalized version of the RH. 
In Connes approach there is an operator, which plays
the role of the Hamiltonian, whose spectrum
is a continuum with missing spectral lines
corresponding to the Riemann zeros. 
We are thus confronted with two possible physical
realizations of the Riemann zeros, either
as point like spectra or as missing spectra
in a continuum. Later on we shall see that 
both pictures can be reconciled in a QM 
model having a discrete spectra embedded in a 
continuum.

\section{Semiclassical approach}

\begin{figure}[t!]
\begin{center}
\includegraphics[height= 9 cm,angle= 0]{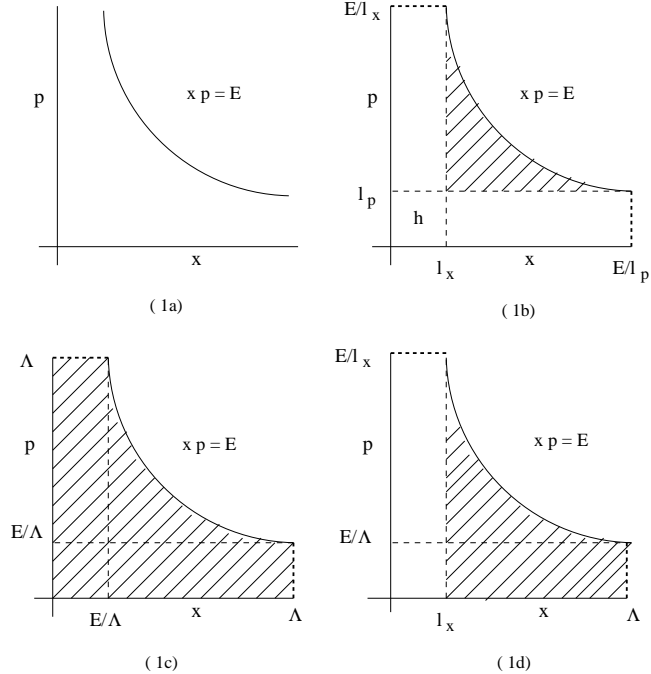}
\end{center}
\caption{
1a) a classical trayectory of the Hamiltonian 
$H = x p$. 
The regions in shadow are the allowed 
phase space of the semiclassical regularizations
of $H = x p$ considered 
by:  1b) Berry and Keating, 1c) Connes and 1d) Sierra. 
}
\label{semiclassical}
\end{figure}

In 1999 Berry and Keating on one
hand \cite{BK1,BK2},  and Connes on the other \cite{Connes}, 
proposed that the
classical Hamiltonian $H = x p$, where $x$
and $p$ are the position and momenta of a 1D particle,
is closely related to the Riemann zeros. 
The classical trayectories 
of the particle are hyperbolas in the phase
space $(x,p)$, hence 
one should not expect a discrete spectrum 
even at the semiclassical level (see fig. \ref{semiclassical}. 
To overcome this difficulty, Berry and Keating proposed
to restrict the phase space 
to those points where  $|x| > l_x$ and $|p| > l_p$, 
with $l_x \, l_p = 2 \pi \hbar$. 
The number of 
semiclassical states, $\CN(E)$,  with energy 
between 0 and $E$ is given by the allowed area
in phase space divided by $h = 2 \pi \; (\hbar =1)$
(see fig. \ref{semiclassical})

\beq
\CN(E) =  \frac{E}{2 \pi} 
\left( \log \frac{E}{2 \pi} -1 \right) + 1
\label{1} 
\eeq

\no
Eq.(\ref{1}) coincides asympotically  with the smooth part of 
the formula that gives the number of Riemann zeros
in the same interval. This result is really striking
given the simplicity of the Hamiltonian and the asumptions made.
On the other hand Connes started on the same classical Hamiltonian
$H= xp$ but constrained the phase space to those
trayectories satisfying $|x| < \Lambda, |p| < \Lambda$,  
with $\Lambda$ a  cutoff which is sent to infinity at the
end of the calculation. The number of semiclassical
states is given  by

\beq
\CN(E) =   \frac{E}{ \pi} \log \Lambda
- \frac{E}{2 \pi} \left( \log \frac{E}{2 \pi} -1 \right)
\label{2}
\eeq

\no
The first term describes a continuum of states
in the limit $\Lambda \rightarrow \infty$, 
while the second term
coincides with minus the average position
of the Riemann zeros (\ref{1}).  This result
lead Connes  to propose the missing spectral
interpretation of the Riemann zeros described
earlier.  

A third possible regularization of the
xp model, proposed in references \cite{Sierra1,Sierra2},
is that $l_x < x < \Lambda$, which leads to the
following counting of semiclassical states,

\beq
\CN(E) =   \frac{E}{ 2 \pi} \log \frac{\Lambda}{l_x} 
\label{3}
\eeq

\no
This result agrees with the asymptotic part of 
(\ref{2}), 
meaning that there is a continuum spectrum but
the possible connection with the Riemann zeros
is lost. The main advantage of the latter
regularization is that it arises
from a consistent quantization of $H = xp$
unlike the two previous semiclassical regularizations.

\section{Quantization of $ x \, p$ and $1/( x p)$}

The classical Hamiltonian $H = xp$
can be consistently quantized in two cases
depending on the choice of the domain in the $x$
coordinate: 1) $0 < x < \infty$ and 2)  $l_x < x < \Lambda$
\cite{Sierra2,Twamley}. In the first case $H$ is essentially self-adjoint, 
while in the second it admits a one parameter 
self-adjoint extension. We shall consider the latter
case. To do so one first define the normal ordered
operator \cite{BK1}

\beq
H_0 = \frac{1}{2} ( x p + p x) = -i (\frac{d}{dx} + \frac{1}{2}) 
\label{4}
\eeq

\no
where $p = - i d/dx$. The formal eigenfunctions of (\ref{4})
are

\beq
\psi_E(x) = \frac{C}{x^{1/2 - i E}}, \qquad 1 < x < N  
\label{5}
\eeq

\no
where we have normalized $l_x=1$ and $\Lambda = N$. 
One can show that (\ref{4}) is self-adjoint 
if the wave functions satisfy the boundary condition

\beq
e^{i \theta} \psi_E(1) =  N^{1/2} \; \psi_E(N)
\label{6}
\eeq

\no
where the angle $\theta$ parameterizes the 
self-adjoint extension of $H$. Imposing (\ref{6})
on (\ref{5}) yields

\beq
  N^{i E }= e^{ i \; \theta}
\label{7}
\eeq

\no
which determines the eigenvalues of $H$

\beq
E_n = \frac{2 \pi}{\log N} \left( n + \frac{ \theta}{2 \pi} \right),
\qquad n = 0, \pm 1, \dots 
\label{8}
\eeq

\no
This equation agrees with the semiclassiclassical
result (\ref{3}). In the particular case where $\theta = \pi$,
the spectrum (\ref{8}) becomes symmetric around 0. 
Notice that the zero eigenvalue is excluded. 
Another way to derive this result is by considering the
inverse of the operator (\ref{4}). This can be done
as follows,

\beq
H_0 = \frac{1}{2} ( x p + p x) = x^\frac{1}{2} \; p \; x^\frac{1}{2}
\rightarrow H^{-1} = x^{-\frac{1}{2}} \; p^{-1} \; x^{-\frac{1}{2}}
\label{9}
\eeq

\no
where $x^{1/2}$ is well defined since $x > 1$. 
The inverse of the momenta operator $p$ is given by the
1D Green function

\beq
p^{-1} = \frac{i}{2} \frac{ \sign(x-x')}{ \sqrt{x \,  x'}},
\label{10}
\eeq

\no
where  $\sign(x-x')$ is the sign function. The Schroedinger
equation associated to (\ref{9}) is

\beq
\frac{i}{2} \int_1^N dx' \;  \frac{ \sign(x-x')}{ \sqrt{x \,  x'}}
\psi_E(x') = \frac{1}{E} \;  \psi_E(x). 
\label{11}
\eeq

\no
which in terms of the new wave function

\beq
\phi_E(x) = x^{-\frac{1}{2}} \; \psi_E(x) 
\label{12}
\eeq

\no
becomes

\beq
x \; \phi_E(x) -
\frac{i \; E }{2} \int_1^N dx' \;  \sign(x-x')
\phi_E(x') = 0 
\label{13}
\eeq

\no
whose solution is

\beq
N^{i E} = -1 , \;\; 
\phi(x) = \frac{C}{ x^{1 - i E}}
\label{14}
\eeq

\no
Hence we recover eqs.(\ref{5}) and (\ref{7})
in the particular case where $\theta = \pi$. 
On the other hand, eq.(\ref{13}) 
looks as the eigenvalue equation
of yet another Hamiltonian that we shall discuss next.

\section{Relation with the Russian Doll BCS model}

In reference
\cite{RD1,RD2,links} it was defined an extension
of the BCS model of superconductivity, called
the Russian Doll model, whose Hamiltonian, 
when restricted to the one body case becomes

\beq
H_{RD}(x,x') = \
\vep(x)  \; \delta(x-x') - \frac{1}{2} 
\left( g + i \; h \; \sign(x-x') \right) 
\label{15}
\eeq

\no
where $\vep(x)$ represents the energies of pairs
of electrons occupying time reversed states 
in the band $1 < x < N$, $g$ is the 
standard BCS coupling constant and $h$ a coupling
that breaks the time reversal symmetry. The eigentates 
and eigenfunctions of (\ref{15}) are given
by

\beq
\left( 
\frac{ N - E_{RD}}{ 1 - E_{RD}} \right)^{i h}
= \frac{g + i h}{g - i h}, \qquad 
\phi(x) = \frac{C}{ (x - E_{RD})^{1 - i h}}
\label{16}
\eeq

\no
Comparing (\ref{16}) with (\ref{14}), one obtains the following map
between the eigenstates of the 
xp model and the RD model

\barray 
& E_{RD} = 0  
\leftrightarrow 
E \neq 0  & 
\label{17}
\\
& h \leftrightarrow E & \nonumber \\
& \frac{h}{g} \leftrightarrow \tan(\theta/2) &
\nonumber \\
& \phi  \leftrightarrow x^{-1/2} \psi_E 
& 
\nonumber 
\earray

\no
in the particular case where $g=0$ and then
$\theta = \pi$. One can add a $g$ coupling term
in the definition (\ref{9}) of 
the operator $H^{-1}$, in which case the
correspondence between the RD model and the 
xp model will cover all the self-adjoint extensions
of $xp$. The RD model provides an example
where the renormalization group, instead of
ending at fixed points, run in cycles \cite{GW,BLflow,fewbody}. 
In the RD model,  
the coupling that runs periodically under the RG 
is $g$, with a period equal to $2 \pi/h$, 
while the coupling $h$ remains invariant. This fact 
in turn  implies the existence of several 
bound states whose number is given by
$n = 2 \pi/h  \log N$. If we replace $h$ by $E$,
the latter equation becomes $n = 2 \pi/E \log N$ 
which coincides with the number of eigentates
of the xp model. 

Incidentally, we would like to mention that
the field theory realizations of the cyclic Renormalization Group, 
of references \cite{LRS1,LRS2,LS},
are at the origin of LeClair's approach to the RH \cite{Andre-RH}.
In this reference the zeta function on the critical strip
is related to the quantum statistical mechanics of non-relativistic,
interacting fermionic gases in 1d with a quasi-periodic 
two-body potential. This quasi-periodicity is reminiscent
of the zero temperature 
cyclic RG of the quantum mechanical Hamiltonian of \cite{Sierra2}, 
but the general framework of both works is different.

\section{$H_0 = x \;p$  with interactions: general model}

The previous results establish
an interesting correspondence between two 
apparently different models which also suggests 
a way to add interactions to the xp model. 
Indeed, the interacting term of the RD Hamiltonian
(\ref{15}) that is proportional to the coupling constant
$g$ is basically a proyector operator $| BCS \rangle
\langle BCS|$, with wave function $\langle x|BCS \rangle = 1$
$\forall x$. As we said above, adding that term to the inverse
Hamiltonian (\ref{9}) would give rise to a $\theta \neq \pi $ term
associated to the self-adjoint extensions of xp. Instead
we want to add an interacting term that reflects the existence
of two boundaries in the BK regularization of the xp model. 
The simplest possibility is to define

\beq
H^{-1} = H_0^{-1} + \frac{i}{2} \left(
|\psi_a \rangle \langle \psi_b | - |\psi_b \rangle \langle \psi_a |
\right)
\label{18}
\eeq

\no
where $\psi_{a,b}$ are two wave functions
whose properties will be specified below. 
The matrix elements of (\ref{18}) read

\beq
H^{-1}(x,x') = \frac{i}{2} 
\frac{ \sign(x-x') + a(x) b(x')  -  a(x') b(x) }{ \sqrt{x \,  x'}},
1 < x, x' < N 
\label{19} 
\eeq

\no
where

\beq
\psi_a(x) = \frac{a(x)}{x^{1/2}}, \;\; \psi_b(x) = \frac{b(x)}{x^{1/2}}
\label{20} 
\eeq

\no
are real functions, which guarantee that $H^{-1}$ is
a hermitean and antisymmetric matrix, so that
its eigenvalues are pairs of real numbers $E, -E \neq 0$. 
A simplified version of (\ref{19}) is obtained
by choosing $b (x) = 1$. The latter model
will be denoted as type I, while the former
as type II. For these models to be well defined
in the limit $N \rightarrow \infty$ we impose
the following normalization conditions

\beq
|\int_1^\infty dx \; \frac{f(x)^n}{x}| < \infty, \;  (n=1,2), \;  
\label{21} 
\eeq

\beq
f = \left\{ \begin{array}{ll}
a  & {\rm type }  \; {\rm I} \\
a, b &   {\rm type }  \; {\rm II} \\
\end{array}
\right. 
\nonumber  
\eeq

\no 
A nice feature of the Hamiltonian (\ref{19})
is that the Scroedinger equation is exactly solvable,
the reason being that it is equivalent to a first
order differential equation supplemented with a boundary
condition. We shall next present the results obtained
in reference \cite{Sierra2}. First of all, the eigenenergies
$E$ of (\ref{19}) satisfy the equation

\beq
\F_N(E) +  \F_N(-E) \; N^{i E}  = 0
\label{22} 
\eeq

\no
where $\F_N(E)$ is a Jost like function
whose expression will given below in the limit $N \rightarrow \infty$. 
In that limit the eigenfunctions of
the type II model are given by

\beq
\psi_E(x) = \frac{1}{x^{1/2 - i E}} 
\left[ C_\infty + \int_x^\infty dy 
y^{-i E} \left(
\frac{d a}{d y} B - \frac{d b}{d y} A
\right) \right] 
\label{23} 
\eeq

\no
where $A, B, C_\infty$ are integration constants
that depend on $E$. In the limit $x \rightarrow \infty$
the functions $a(x), b(x)$ vanish sufficiently
fast so that 
the wave function (\ref{23})
is dominated by the first term, i.e.

\beq
\lim_{ x >> 1} \psi_E(x) \sim \frac{C_\infty}{x^{1/2 - i E}} 
\label{24} 
\eeq

\no
It turns out that 
$C_\infty$ is given by the Jost function (up to a constant
which can be taken as 1),

\beq
C_\infty(E) =   \F(E) 
\label{25} 
\eeq

\begin{figure}[t!]
\begin{center}
\includegraphics[height= 2.5 cm,angle= 0]{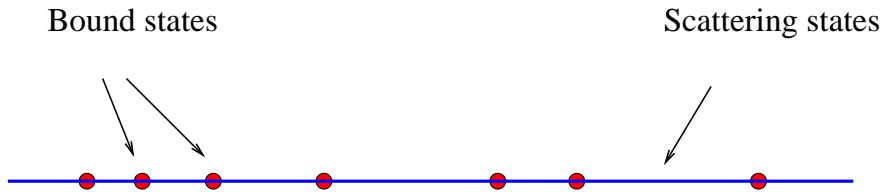}
\end{center}
\caption{
Pictorial representation of the spectrum of the model.
The bound states are the points where $\F(E) = 0$, which
are embedded in a continuum of scattering states. 
}
\label{bound-states}
\end{figure}

\no
Hencefore the energies where $\F(E)$ is non zero
correspond to delocalized states which behave
asymptotically as the eigenfunctions of 
the unperturbed Hamiltonian $H_0 = ( x p + p x)/2$.
For these states the ratio $\F(E)/\F(-E)$
gives the scattering phase shift. On the other hand,
 $C_\infty$ vanishes whenever $\F(E)$
 does. These energies correspond to localized states
with a finite norm. In summary, the spectrum of $H$
consists of a continuum formed by those energies
where $\F(E) \neq 0$, plus a discrete part given
by the real zeros of $\F(E)$ (see fig. \ref{bound-states}).

Moreover, using the hermiticity
of $H$ one can show from eq.(\ref{21})
that $\F_\infty(E)$ does not have zeros
with $\Im (E) > 0$.

\beq
\CF(E) =0 \Rightarrow \Im \; E \leq 0 
\label{26} 
\eeq

\no 
The  real zeros of $\CF(E)$ correspond, as
explained above, to localized states, while the complex zeros
below the real axis correspond to resonances.
These results are summarized in table 1.

\begin{center}
\begin{tabular}{|c|c|c|c|}
\hline
Eigenstate & $C_\infty$  &  $\F(E)$ &  Eigencondition \\
\hline 
Delocalized & $\neq 0$ & $\neq 0$ & $N^{i E} = -
\frac{\F(E)}{\F(E)^*} $ \\
Localized & $= 0$ & $ = 0$ & $\F(E) = 0$ \\
\hline
\end{tabular}

\vspace{0.5 cm}

Table 1.- Classification of eigenstates of the model.
\end{center}

Before giving the expression of $\F(E)$ for the type I and type II
model 
we shall introduce some definitions. 
First of all let us define the Mellin transform of $a$ (similarly
for $b(x)$).

\beq
\ha(t) = \int_{1}^\infty dx \; x^{-1 + i t} \,a(x),   \quad
 a(x) = \int_{- \infty}^\infty \frac{dt}{2 \pi} \, x^{i t} \;  \ha (t) 
\label{27}
\eeq

\no
The reality of $a(x)$ implies

\beq
\ha^*(t) = \ha(-t), \;\; t \in \Rmath 
\label{28} 
\eeq

\no
Condition $n=1$ in eq.(\ref{21}) amounts to

\beq
|\ha(0)| < \infty 
\label{29} 
\eeq

\no
while condition  $n=2$ in eq.(\ref{21}) is equivalent to

\beq
\int_1^\infty dx \;  \frac{a(x)^2}{x} = 
\int_{-\infty}^\infty \frac{dt}{2 \pi} \;  | \ha(t) |^2
 < \infty,
\label{30} 
\eeq

\no
The function $\ha(t)$ is in fact the Fourier transform
of $a(x)$ in the variable $q= \log x$, which takes values
in the interval $(0,\infty)$. In terms of $q$, 
$a(x)$ is a square normalizable and 
causal function, in which case $\ha(t)$ 
has interesting analytic properties 
by  a theorem due to Titchmarsh \cite{Titchmarsh}. This theorem states
that under the previous conditions 
$\ha(t)$ is analytic in the complex upper-half plane
and satisfies the formula

\beq
\ha(z) = P \int_{- \infty}^\infty \frac{dt }{ i \pi}
\frac{ \ha(t)}{t - z}, \quad z \in \Rmath 
\label{31} 
\eeq

\no
where P denotes the Cauchy principal value of the integral. 
To prove this formula one uses the fact that
$\ha(t)$ has no singularities in the upper-half plane
and that the contour on integration on the circle
$|z| = R, \Im \;  z > 0$ vanishes since $\lim_{|z| \rightarrow
\infty} |\ha(z)| = 0$. For later purposes let us define 
the new function

\beq
\a(t) = \frac{i t}{2}  \ha(t),
\label{32} 
\eeq

\no
and similarly $\b(t)$, whose properties follow
from those of $\ha(t)$ namely:

\begin{itemize}

\item Reality:

\beq
\a^*(t) = \a(-t), \;\; t \in \Rmath 
\label{33} 
\eeq

\item Regularity:

\beq
\lim_{ t \rightarrow 0} 
\frac{ |\a(t)|}{t} < \infty \Longrightarrow 
|\a(0)| = 0 
\label{34} 
\eeq

\item Normalizability

\beq
\int_{-\infty}^\infty \frac{dt}{2 \pi} \; \frac{|\a(t)|^2}{t^2} 
 < \infty,
\label{35} 
\eeq

\item Analiticity

\beq
\a(z) = P \int_{- \infty}^\infty \frac{dt }{ i \pi}
\frac{ \a(t)}{t - z} - 
 P \int_{- \infty}^\infty \frac{dt }{ i \pi}
\frac{ \a(t)}{t}, 
\quad z \in \Rmath 
\label{36} 
\eeq

\no
This eq. implies that $\a(z)$ is an analytic function
in the upper-half plane which converges towards a constant
value in a circle of infinite radius  given 
by the last term of (\ref{36}).

\end{itemize}

The next definition involves the product of 
two analytic functions $\f(t)$ and 
$\g(t)$ in the upper half-plane:

\beq
(\f \star \g)(z) = \f(z) \;  \g(-z) + 
\int_{- \infty}^\infty \frac{dt }{ i \pi}
\frac{ \f(t) \;  \g(-z)}{t - z} , \quad z \in \Rmath
\label{37} 
\eeq

\no
where the integration is understood in the Cauchy sense 
as in eqs.(\ref{31}) and (\ref{36}). 
One can show
that $(\f \star \g)(z) $ is an analytic function in the
upper half-plane provided  $\f(z) \;  \g(-z)$ is well
behaved, which seems to be the case in all the examples
we have analized. The analytic extension of $(\f \star \g)(z) $
to the lower half-plane will have in general singularities. 
In terms of this product we shall define the function
\cite{sfg}:

\beq
S_{\f,\g}(z) = (\f \star \g)(z) -  (\f \star \g)(0)
\label{38} 
\eeq

\no
which satisfies the following conditions

\begin{itemize}

\item Reality: if  $\f$ and $\g$ verify (\ref{33}) then

\beq
S_{\f,\g}^*(z) =S_{\f,\g}(-z), \quad z \in \Rmath  
\label{39} 
\eeq

\item Regularity

\beq
S_{\f,\g}(0) = 1
\label{40} 
\eeq

\item Shuffle relation

\beq
S_{\f,\g}(z) +S_{\g,\f}(-z)  = 2  \; \f(z) \; \g(-z)
\label{41} 
\eeq

\no
The notation ``shuffle'' is borrowed from the Theory
of multiple zeta functions, 
as explained later on.

\end{itemize}

\no
After  these definitions we can finally give
the expression of the Jost function $\F(E)$
in terms of the potentials $\a(t)$
and $\b(t)$. For the type I model it reads

\beq
\F(t) = 1 + 2 \;  \a(t) + S_{\a, \a}(t) 
\label{42} 
\eeq

\no
while for the type II  model it is

\beq
\F(t) = 1 -  S_{\a, \b}(t) + S_{\b, \a}(t) +  
 S_{\a, \a}(t) S_{\b, \b}(t) - S_{\a, \b}(t) S_{\b, \a}(t)
\label{43} 
\eeq

\no
From the properties of the $S$-functions one can easily
derived:

\begin{itemize}

\item Reality:

\beq
\F^*(t) = \F(-t)
\quad z \in \Rmath  
\label{44} 
\eeq

\no
This condition guarantees that the ratio $\F(E)/\F(-E)$
appearing in the eigenvalue  eq.(\ref{22}) is indeed a phase.

\item Regularity

\beq
\F(0) = 1
\label{45} 
\eeq

\no
This condition implies that the zero energy
is not an eigenvalue of the Hamiltonian $H$,
which was the assumptions made by defining it
in terms of its inverse.

\end{itemize}

Let us next consider the two models separately.

\subsection{Type I model}

The Jost function (\ref{42}) can be expressed as

\beq
\F(t) = 1 + 2 \;  \a(t) + (\a \star \a)(t) 
\label{46} 
\eeq

\no
where we used that $\a(0) = 0$, which implies

\beq
(\a \star \a)(0) = 0.
\label{47} 
\eeq

\no
An important consequence of eq.(\ref{46}) 
is the positivity of the real
part of $ \F(t)$,

\beq
\Re \;  \F(t) = |1 +  \a(t)|^2 \geq 0  
\label{48} 
\eeq

\no
which imposes a strong contraint on the functions
allowing a  QM interpretation as Jost functions of the
type I   model. In particular
eq.(\ref{48}) excludes the zeta function $\zeta(\sigma - i t)$
for $1/2 \leq \sigma < 1$, but not the case where
$\sigma > 1$ as we shall see later on. 

The $\star$-product defined in eq.(\ref{37}) is non commutative
and non associative. Nevertheless it behaves nicely respect
to the identity function $1$, namely 

\beq
1 \star 1 = 1 
\label{49}
\eeq

\no
and

\beq
\a \star 1 + 1 \star \a  = 2 \; \a 
\label{50}
\eeq

\no
where $\a$ is an analytic function in the upper half-plane
satisfying eq.(\ref{36}). Using these two equations, 
the Jost function (\ref{46}) can be expressed as 
the {\em square}, with respect to the $\star$-product,  
 of a single function, i.e.

\beq
\F = \f \star \f, \qquad \f = 1 + \a
\label{51} 
\eeq

\no
so that $\f$ is the $\star$-{\em square  root}, 
of $\F$. 
Using eq.(\ref{51}) one can easily prove
that $\F(t)$ does not have zeros in the upper half-plane.
Indeed, write  $\f \star \f$  as

\beq
(\f \star \f)(z) = 
\int_{- \infty}^\infty \frac{dt }{ i \pi}
\frac{ \f(t) \;  \f(-t)}{t - z} , \quad \Im \; z > 0 
\label{52} 
\eeq

\no 
then

\beq
\Re (\f \star \f)(z) = 
\int_{- \infty}^\infty \frac{dt }{  \pi}
\frac{ y \; |\f(t)|^2 }{(t - x)^2 + y^2} > 0, \quad z = x + i y  
\label{53} 
\eeq

\no 
A simple but illustrative example of the theory
developped so far is provided by the step
potential,

\beq
a(x) = \; \left\{
\begin{array}{lcl}
a_1, & & 1 < x < x_1 \\
0, & & x_1 < x < \infty \\
\end{array}
\right. 
\label{54}
\eeq

\no
which yields

\beq
\a =   \frac{a_1}{2} \;( x_1^{i t } -1), \qquad  
\a \star \a   =    \frac{a_1^2}{2} \;(1-  x_1^{i  E}), 
\label{55}
\eeq

\no
and the Jost function

\beq
\F(t) = 1 + \frac{a_1 \;( 2 - a_1)}{2}  ( x_1^{iE} -1). 
\label{56}
\eeq

\begin{figure}[t!]
\begin{center}
\includegraphics[height= 6 cm,angle= 0]{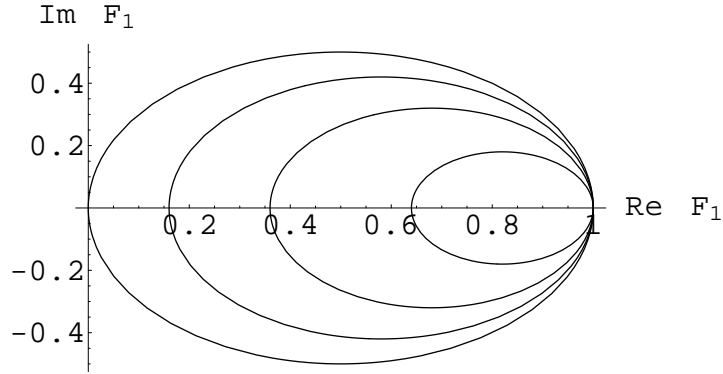}
\end{center}
\caption{
Plot of the real and imaginary parts of $\F(E)$,
as given by eq.(\ref{56}), for the choices
$a_1 = 0.2, 0.4, 0.6, 1$. At $a_1= 1$ the circle
passes through the origin. 
}
\label{step1}
\end{figure}

\no
Fig. \ref{step1}  shows an Argand plane representation
of the real and imaginary parts of (\ref{56}) for
several values of $a_1$. For $a_1 = 1$ 
the function $\F(t)$ vanishes at the values

\beq
E_{n} = \frac{(2 n + 1) \pi}{\log x_1},\; \; n = 0, \pm 1, \dots 
\label{57}
\eeq

\no
describing an infinite number of bound states. The remaining values
of $E$ correspond to delocalized states.
All the zeros of $\F(E)$  lie on the real axis for $a_1 = 1$
and below it for $a_1 \neq 1$.

\subsection{Perturbative solution of the type I  model}

The next problem we address is: given 
a function $\F(t)$, satisfying the analyticity,
reality, regularity and positivity conditions 
described above, which is the potential, or potentials,
$\a(t)$, verifying eq.(\ref{46})?  In this paragraph
we shall give a perturbative method to construct 
one of those potential in terms of a series
which converges under certain conditions placed
on the function $\F(t)$.  
Let us first make the change $\a \rightarrow - 2 \a$
in eq.(\ref{46}) which becomes

\beq
\a = \g + \a \star \a, \qquad \g = \frac{1 - \F}{4} 
\label{58}
\eeq

\no
Iterating (\ref{58}) generates  the series expansion

\beq
\a = \g + \g \star \g  + \g \star(\g \star \g) +  (\g \star \g)
\star \g  + \dots
\label{59}
\eeq

\no
At order $\g^n$ one gets all admissible bracketings,
whose number is given by $C_{n-1}$, where $C_n$
is the Catalan number

\beq
C_n = \frac{1}{n+1} \left( 
\begin{array}{l}
2 n \\
n 
\end{array}
\right)  \stackrel{n >> 1}{\rightarrow} \frac{4^n}{\sqrt{\pi} \; n^{3/2}}
\label{60}
\eeq

\no
whose asymptotic behaviour is also described. 
To investigate the conditions for convergence 
of (\ref{59}), let us suppose 
that $\g$ is given by the absolute convergent series

\beq
\g(t) = \sum_{n=1}^\infty g_n \; n^{i  t}, \quad 
|\g(t)| \leq \sum_{n=1}^\infty |g_n| < \infty 
\label{61}
\eeq

\no
Using  (\ref{37}) one finds

\beq
(\g \star \g)(t) = \sum_{n=1}^\infty g_n^2 + 
2 \sum_{n > m} g_n \, g_m \; (n/m)^{i t} 
\label{62}
\eeq

\no
so that

\beq
|(\g \star \g)(t)| \leq (\sum_{n=1}^\infty |g_n|)^2, \; \forall t \in \Rmath
\label{63}
\eeq

\no
Similarly one finds that

\beq
|\a(t)| \leq 
\sum_{n=1}^\infty  C_{n-1} \;  \left(\sum_{m=1}^\infty |g_m| \right)^n,
\label{64}
\eeq

\no
which, according to (\ref{60}), converges provided

\beq
\sum_{m=1}^\infty |g_m|  \leq \frac{1}{4} 
\label{65}
\eeq

\no
This condition is 
sufficient for convergence of the series (\ref{59})
but it is not necessary. Eq.(\ref{65}) implies

\beq
|\F(t) - 1| \leq 1, \;\; \forall t \in \Rmath 
\label{66}
\eeq

\no
However the converse is not true (i.e. eq.(\ref{66})
does not imply (\ref{65})). 
We believe that (\ref{66}) 
also guarantees the convergence of
(\ref{59}), but this guess needs to be proved. 
Observe also that (\ref{66}) implies that 
$\Re \; \F(t) \geq 0$, which is also
a necessary condition for the existence
of the potential $\a(t)$. 
An application of these results is: 

\subsection{The zeta function with $\sigma > 1$}

For  $\sigma > 1$ let us consider the function

\beq
\F(t) = C \; \zeta(\sigma - i t),  
\quad C = \frac{1}{\zeta(\sigma)}
\label{67}
\eeq

\no
where the constant $C$ guarantees the normalization
condition $\F(0) = 1$. The values of the constants
$g_n$ appearing in (\ref{61}) are given by

\beq
g_1 = \frac{1-C}{4},\quad g_{n>1} = -\frac{C}{4 \; n^\sigma} 
\label{68}
\eeq

\no
and the convergence criteria (\ref{65})  
 yields

\beq
\sum_{m=1}^\infty |g_m| = \frac{1 - C}{2} \leq \frac{1}{4}
\rightarrow \zeta(\sigma) \leq 2  
\label{69}
\eeq

\no
where the latter conditions is satisfied if

\beq
\sigma \geq \sigma_c = 1.72865, \;\; \zeta(\sigma_c) = 2  
\label{70}
\eeq

\no
Moreover condition (\ref{66}) can be checked
numerically  i.e.

\beq
|\frac{\zeta(\sigma - i t)}{\zeta(\sigma)} - 1 | \leq 1, \quad
\forall \sigma > 1, \;\; \forall t \in \Rmath 
\label{71}
\eeq

\begin{figure}[t!]
\begin{center}
\hbox{\includegraphics[height= 7 cm, width= 7 cm]{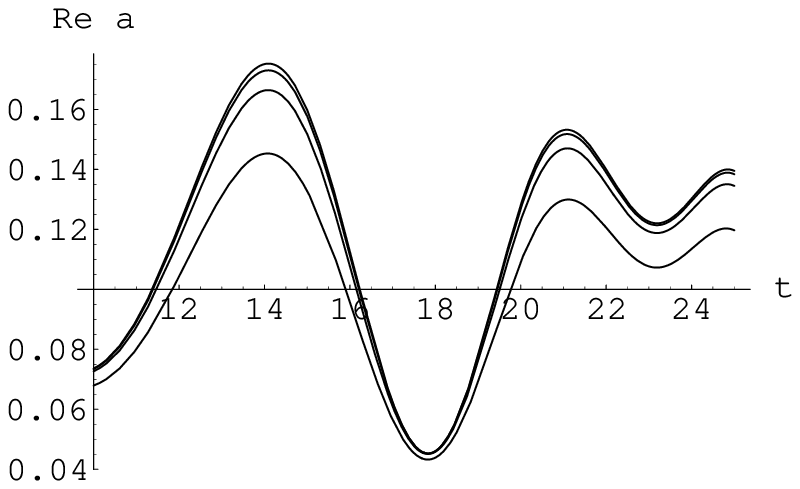}
\includegraphics[height = 7 cm, width= 7 cm]{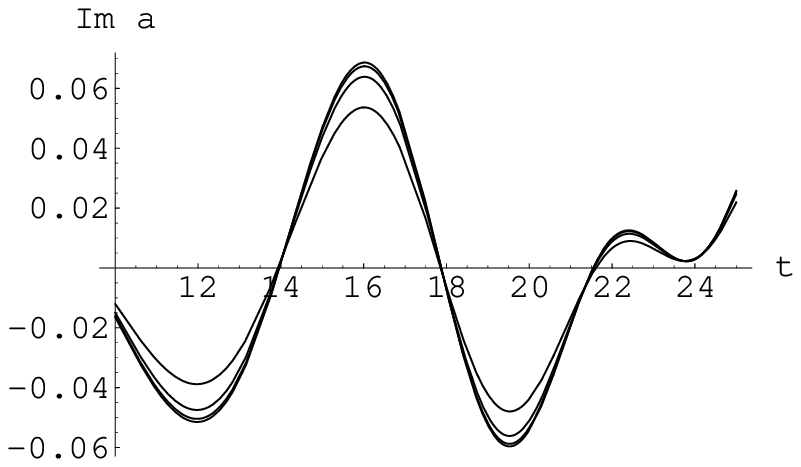}}
\end{center}
\caption{Real and imaginary parts of the potential
$\a(t)$ for $\sigma = 2$ in the interval 
$t \in (10,25)$. 
}
\label{Re-Im-s2}
\end{figure}

\no
so we expect that the series (\ref{59}) will 
also converge for any value $\sigma > 1$.
Fig. \ref{Re-Im-s2} displays the real
and imaginary parts of $\a(t)$ for the
case $\sigma = 2$ obtained by the sum 
of the first terms of eq.(\ref{59}). 
The convergence towards a finite value is clear.

The series (\ref{59}) contains also an interesting analytical
structure, which can be seen from the 
star product of two zeta functions,

\beq
\zeta(\sigma - i t) = \sum_{n=1}^\infty \frac{1}{n^{\sigma - i t}}
\rightarrow (\zeta \star \zeta)(t) =
\zeta(2 \sigma) + 2 \sum_{n> m \geq 1}^\infty 
 \frac{1}{n^{\sigma - i t} \; m^{\sigma + i t}} 
\label{71-1}
\eeq

\no
The double sum series of the RHS is equal to 
a Euler-Zagier zeta function for two variables
which is a generalization of the zeta
function. Multivariable versions of this
function  
have attracted much attention 
in various fields, as  knot theory,
perturbative quantum field theory, etc
(see \cite{euler-zagier-1},  \cite{euler-zagier-2}
and references therein).

\subsection{The zeta function at $\sigma = 1/2$}

\begin{figure}[t!]
\begin{center}
\hbox{\includegraphics[height= 6.5 cm, width= 7.9 cm]{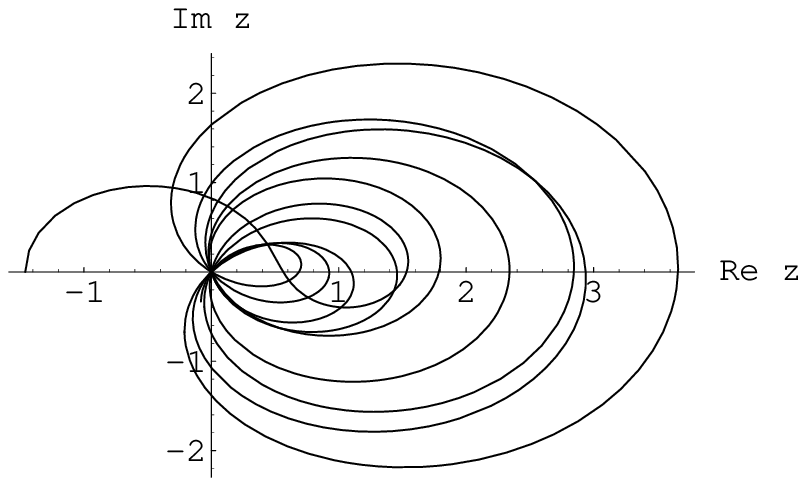}
\includegraphics[height = 6.5 cm, width= 6.5 cm]{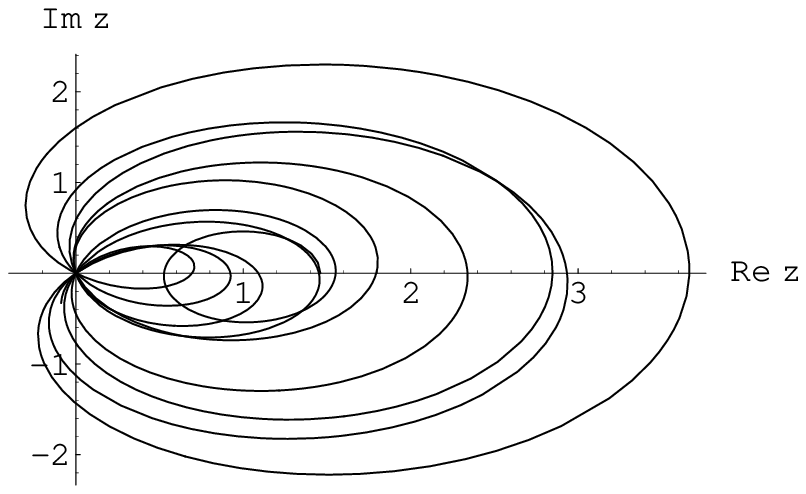}}
\end{center}
\caption{Left: real and imaginary parts of $\zeta(s)$
with $s= 1/2 - i E$ and $E \in (0, 50)$.
Right: same as before for  $\zeta_H(s) = (s-1) \zeta(s)/s$. 
}
\label{zeta1}
\end{figure}

The results obtained so far suggests that the 
Riemann zeta function on the critical line
$\zeta(1/2 - i E)$ could perhaps be realized
as the Jost function of the model. This idea 
is motivated by 
the  scattering approach pionered by Faddeev and Pavlov
in 1975, and has been followed by many authors 
\cite{Faddeev,Lax,G2,Joffily}. An important result
is that the phase of $\zeta(1 + i t )$ is related
to the scattering phase shift of a particle moving 
on a surface with constant negative curvature. The chaotic nature of that
phase is a well known feature. Along this
line of thoughts, Bhaduri, Khare and Law (BKL) maded in 1994 
an analogy between  resonant quantum scattering amplitudes 
and the Argand diagram of the zeta function
$\zeta(1/2 - i t)$, where
the real part of $\zeta$ (along the $x$-axis) 
is plotted against the imaginary part ($y$-axis) \cite{BKL}. 
The diagram consists of an infinite series of closed loops
passing through the origin every time $\zeta(1/2 - i t)$
vanishes (see fig. \ref{zeta1}). This loop structure
is similar to the Argand plots of  partial wave amplitudes
of some physical models with the two axis being interchanged.
However the analogy is flawed since the real part of $\zeta(1/2  - it)$
is negative in small regions of $t$, a circumstance
which never occurs in those physical systems.

In fact, the loop structure of the models proposed by BKL
is identical, up to a scale factor of 2, to the model
with the potential (\ref{54}) 
(see fig. (\ref{step1})), where 
the loops representing $\F(E)$, for $a_1 = 1$,  
are circles of radius 1/2, centered
at $x=1/2$. For general models of type I, the loops
are not circles but 
the real part of $\F_1(E)$ is always positive (see eq.(\ref{48})), 
and therefore they can never represent $\zeta(1/2 - i E)$.
Incidentally, this constraint does not apply to the models
of type II, where $\Re \; \F(E)$ may become negative.
This suggests that
$\zeta(1/2 - i E)$ could indeed be the Jost
function $\F(E)$ of a type II model for a particular choice
of $a$ and $b$. 
If this were
the case then the Riemann zeros would become
eigenenergies of the Hamiltonian realizing in that
manner the Polya and Hilbert conjecture which
may also give hints into the solution of the Riemann
hypothesis. A complete answer to this problem is not yet known 
but we shall present below some encouraging results along this
direction.

The first step is to recover quantum mechanically
the smooth approximation to the Riemann zeros. 
This approximation is equivalent to the following 
condition

\beq
1 + e^{ 2 i \theta(E)} = 0
\label{72}
\eeq

\no
where

\beq
e^{ 2 i \theta(E)} = \pi^{-i E} \frac{ 
\Gamma \left( \frac{1}{4} + \frac{i E}{2} \right)}
{ \Gamma \left( \frac{1}{4} - \frac{i E}{2} \right)}
\label{73}
\eeq

\no 
The function $\theta(E)$ gives  the phase of the zeta function, e.g.

\beq
\zeta(1/2 - i t) = Z(t) \; e^{i \theta(t)} 
\label{74}
\eeq

\no
while $Z(t)$ is the  Riemann-Siegel zeta function
which is real and even due to the duality relation
satisfied by the zeta function (e.g. $\zeta(1/2 - i t)
= e^{2 i \theta(t)} \; \zeta(1/2 + i t)$). 
The reason that  (\ref{72}) is a good approximation
to the location of the zeros, 
can be seen in fig. \ref{zeta1} which plots the real and imaginary parts
of $\zeta(1/2 - i E)$. Observe that in the vicinity
of a zero, the curves cut the imaginary axis where 
$\theta(E_n) = \pi (2 n -1)/2$
so that $\cos \theta(E) = 0$, which is nothing but 
eq.(\ref{72}). The value of $E_n$ satisfies

\beq
n = \frac{\theta(E_n)}{\pi} + \frac{1}{2} 
\label{75}
\eeq

\no
which asymptotically coincides with eq. (\ref{1})
up to a constant term. This result can be obtained
choosing the following potential in the type I
model

\beq
a(x) = - 2 \; \frac{ \sin( 2 \pi x)}{\sqrt{x}} 
\label{76}
\eeq

\no
whose Mellin transform (\ref{27}) gives

\beq
\ha(t) = - \frac{2 i}{t} e^{2 i \theta(t)} + O(1/t) 
\label{77}
\eeq

\no
up to $1/t$ terms. The corresponding Jost function
is

\beq
\CF(E) = 2 ( 1 + e^{2 i \theta(E)} ) + O(1/E) 
\label{78}
\eeq

\no
whose zeros agrees with (\ref{72}) asymptotically. 
Hence the smooth approximation to the Riemann zeros
can be obtained asymptotically by the potential (\ref{76}).
Why this potential is able to yield this result? 
One may suspect that it must implement  
at the quantum level the BK  boundary
conditions of the semiclassical approach. 
Indeed, let us show how this works in detail. 
The potential (\ref{76}) corresponds to the wave function

\beq
\psi_a(x) = - 2 \; \frac{ \sin( 2 \pi x)}{x} 
\label{79}
\eeq

\no
which satisfies the equation

\beq
H_0^2  \;  \psi_a (x) = \left( (2 \pi x )^2  - \frac{1}{4} \right)
  \psi_a(x), 
\label{80}
\eeq

\no
where  $H_0 = \sqrt{x} p \sqrt{x}$ is the Hamiltonian
(\ref{9}).  Dropping  $\psi_a$ in both sides
and replacing $H_0$ by $ xp$, one obtains a 
 classical version of  (\ref{80}),

\beq
(x p )^2 =  ( 2 \pi x)^2  - \frac{1}{4} \Longrightarrow
p = \pm  2 \pi  \sqrt{ 1 - \frac{1}{(4 \pi  x)^2}},  
\label{81}
\eeq

\no
which describes a curve in  phase space that approaches asymptotically
the lines $p = \pm 2 \pi$. 
We shall identify
these  asymptotes with the BK boundary  
in the momenta $|p| = l_p$.
Recall on the other hand  the
boundary condition $ x \geq l_x = 1$, which 
combined with the previous identification 
reproduces the Planck cell quantization condition,

\beq
l_p = 2 \pi, \;\; l_x = 1
\Longrightarrow l_p \,  l_x = 2 \pi 
\label{82}
\eeq

\no
The BK condition $|x| > l_x $ has already 
built in into the model and this is also
reflected in the state $\psi_b(x) = 1/\sqrt{x}$
which is concentrated near the position $l_x = 1$.
In the model we have proposed, the two BK boundary
conditions are realized asymmetrically, as opposed
to the semiclassical model. It would be desirable
to have a more symmetric treament of them. This indeed
can be done and the results will be presented elsewhere. 

The next natural step is to see wether 
the zeta function $\zeta(1/2 - i E)$
can be realized as the Jost function of the model.
The analyticity properties of the Jost functions
implies that $\CF(E)$ must be of the form

\beq
\CF(E) = C \frac{E - i/2}{E + i \mu} \; \zeta(1/2 - i E)
\label{83}
\eeq

\no
which does not have poles in the upper-half plane
($C$ and $\mu > 0$ are normalization constants).
In  fig.  \ref{zeta1} we plot an example with $\mu = 1/2$. 
Since the real part of  $\zeta(1/2 - i E)$, and thus
of $\CF(E)$, become negative in small regions of $E$
one is forced to consider the type II model
with two non trivial functions $\psi_a$ and $\psi_b$. 
The problem of finding $\psi_{a,b}$ is rather non trivial.
One can in principle fix one of them, say $\psi_b$,  
and try to solve for $\psi_a$ as a function of the 
Jost function (\ref{83}) and $\psi_b$. 
In the case of the 
zeta function $\zeta(\sigma - i E)$, with 
$\sigma > 1$, we were able
to solve this problem perturbatively thanks
to the fact that the zeta
function is bounded. However for $\sigma =1/2$
the zeta function is unbounded which may lead to 
problems of convergence. In any case, it seems clear
that one needs further physical insights to
make progress into  this difficult problem.  
As we said above, one needs a more symmetric
treatment of the coordinates $x$ and $p$, and a clearer
physical interpretation of the wave functions
$\psi_{a,b}$. Another important ingredient 
to be implemented is  the duality symmetry
of the zeta function which in the present
formulation of the model is not manifest
but which is expected to play a central role. 

In summary, we have presented in this work
an interacting version of the xp Hamiltonian
which may ultimately lead to a spectral realization
of the Riemann zeros, as suggested long ago by Polya and Hilbert. 
The main ingredient of the model is the non local
character of the interaction in terms of two potentials
 which are the quantum 
version of the semiclassical phase space constraints 
of Berry and Keating. The generic spectrum of the model
consists of a continuum of eigenstates in the thermodynamic
limit which may also contain bound states
embedded in it. We conjecture the existence of potentials
giving rise to the Riemann zeros as the discrete spectrum
embedded in the continuum. If this were the case
that would resolve the emission versus absortion
spectral interpretation of the Riemann zeros. 
This would also open the way to a better understanding of the
Riemann hypothesis. We have also pointed out the
need to implement in an explicit way the duality
properties of the zeta function, which implies
a more symmetric treatment of the $x$ and $p$
coordinates as in the semiclassical model.

\section{Acknowledgments}
I would like to thanks the organizers of the 
``5th International Symposium on Quantum Theory and Symmetries''
held at University of Valladolid, Spain, 
and specially Mariano del Olmo, for the invitation to present
the results of the present work. 
I wish also to thank Andre LeClair 
for the many discussions we had on our joint work on the 
Russian doll Renormalization Group 
and its relation to the Riemann hypothesis. 
I also thank M. Asorey, M. Berry, 
L.J. Boya,  J. Garc\'{\i}a-Esteve, J. Keating,  
M.A. Mart\'{\i}n-Delgado, G. Mussardo, 
J. Rodr\'{\i}guez-Laguna and J. Links for our conversations.  
This work was supported by the 
CICYT of Spain
under the contract FIS2004-04885.  I also acknowledge 
ESF Science Programme INSTANS 2005-2010.

\end{document}